\documentclass[aps,prl,twocolumn,superscriptaddress,showpacs]{revtex4-1}
\usepackage{graphicx}% Include figure files
\usepackage{dcolumn}% Align table columns on decimal point
\usepackage{bm}% bold math
\usepackage{subfigure}
\usepackage{amsmath}
\usepackage{hyperref}

\newcommand{\ua}{\uparrow}
\newcommand{\da}{\downarrow}
\newcommand{\ra}{\rightarrow}

\newcommand{\bs}{\boldsymbol}

\newcommand{\SRO}{Sr$_2$RuO$_4$}

\usepackage{times}

\begin{document}
\title{Superconducting pairing in \SRO~from weak to intermediate coupling}
\author{Li-Da Zhang}
\address{School of Physics, Beijing Institute of Technology, Beijing 100081, China}

\author{Wen Huang}
\email{huangw001@mail.tsinghua.edu.cn}
 \address{Institute for Advanced Study, Tsinghua University, Beijing 100084, China}

\author{Fan Yang}
\email{yangfan\_blg@bit.edu.cn}
\address{School of Physics, Beijing Institute of Technology, Beijing 100081, China}

\author{Hong Yao}
\email{yaohong@tsinghua.edu.cn}
\affiliation{Institute for Advanced Study, Tsinghua University, Beijing 100084, China}
\date{\today}
\affiliation{State Key Laboratory of Low Dimensional Quantum Physics, Tsinghua
University, Beijing 100084, China}

\begin{abstract}
The unconventional superconductivity in \SRO~continues to attract considerable interest. While many measurements can be interpreted on the basis of chiral $p$-wave pairing with intriguing topological character, a number of exceptions hinder an unambiguous verification of such pairing. The pairing mechanism also remains under debate. In this paper, with effects of the sizable spin-orbit coupling accounted for, we reexamine the superconducting instabilities in \SRO~through systematic microscopic analysis within random phase approximation. Our calculations show that the odd-parity $p$-wave pairing is favored in the regime of extremely weak interactions, but that highly anisotropic even-parity pairings become most leading over a broad range of stronger interactions. These results could shed light on the nature of the enigmatic superconductivity in \SRO.
\end{abstract}

\maketitle
The superconductivity in \SRO~has attracted significant attention since it was discovered in 1994 \cite{Maeno:94,Maeno:01,Mackenzie:03}. The discovery was followed by proposals of $p$-wave pairing arising from possible remnant ferromagnetic correlations in the material \cite{Rice:95,Baskaran:96,Cuoco:98}, which was later substantiated by experimental evidences of spin-triplet and odd-parity pairing \cite{Ishida:98,Duffy:00,Nelson:04,Ishida:15,Manago:16}. The observation of time-reversal symmetry breaking \cite{Luke:98,Xia:06} raised the prospect of topological chiral $p$-wave pairing which supports Majorana fermions \cite{Kopnin:91,Read:00}. However, despite many progresses, the chiral $p$-wave order has not been fully established, largely due to the difficulties in reconciling it with a variety of other measurements \cite{Kallin:09,Maeno:12,Kallin:12,Mackenzie:17}. We highlight some major issues below. 

A major prediction for chiral $p$-wave is the appearance of spontaneous chiral currents at the edges and domain walls separating regions of opposite chiralities \cite{Matsumoto:99,Furusaki:01}. However, no clear signature of chiral current have been detected so far \cite{Kirtley:07,Hicks:10,Curran:14}. We stress that, as the current is not topologically protected \cite{Huang:15}, it can be suppressed, to various degrees, by gap and band structure anisotropy, surface disorder and the multiband effects \cite{Ashby:09,Imai:12,Bouhon:14,Lederer:14,Scaffidi:15,Huang:15,Etter:17}. However, acquiring full consistency with the experimental null result in the best prepared sample/device is perhaps only possible with a fine-tuned gap structure. Remarkably, edge current could vanish in non-$p$-wave chiral superconductors such as chiral $d$- and $f$-wave, etc \cite{Huang:14,Tada:15,Ojanen:16,Suzuki:16}. Yet these states may be less likely in \SRO~\cite{Huang:14}.

Another discrepancy is the absence of split transitions when an in-plane magnetic field or a uniaxial strain breaks the degeneracy between the two chiral components \cite{Sigrist:91,Yonezawa:14,Hicks:14,Steppke:17,Liu:17,Hsu:16,Liu:13}. A further puzzle is the anomalous suppression of the in-plane upper critical field $H_{c2}$ \cite{Deguchi:02} at low temperatures and a related first-order nature of the superconducting transition \cite{Yonezawa:13,Yonezawa:14,Kittaka:14}. The suppression of the in-plane $H_{c2}$ was also implicated in the vortex lattice anisotropy throughout the mixed state \cite{Rastovski:13,Kuhn:17}. These resemble the Pauli limiting behavior, and can usually be tied to spin-singlet pairing, but may also be explained if the ${\bs d}$-vector of the spin-triplet pairing lies in the plane  -- which however corresponds to a time-reversal invariant helical state. Equally intriguing is the evidence of line nodes \cite{Ishida:97,NishiZaki:00,Hassinger:16}, whereas chiral $p$-wave is typically fully-gapped except in special cases where anisotropy introduces accidental nodes or deep gap minima \cite{Kivelson:13}.

Part of the complication arises from the multiband nature \cite{Deguchi:04}, which hinders unambiguous interpretations of some of the experiments. Since early times it was pointed out \cite{Agterberg:97} that superconductivity is most likely dominated by one set of the three metallic bands, given the disparity of the quasi-1D and quasi-2D bands. By interband scattering, pairing can also develop on the subdominant band(s) with a weaker amplitude \cite{Zhitomirsky:01}. However, there seems to be no consensus regarding the identification of the primary superconducting band(s). The van Hove singularity on the $\gamma$-band and the wall of enhanced spin fluctuations between $\bs Q_1 \approx (2/3,2/3)\pi$ and $\bs Q_2 \approx (\pi,2k_F)$ associated with the quasi-1D bands \cite{Imai:98,Braden:02,Mazin:99} were argued to promote independent pairing instabilities, including $p$-wave, on the respective band(s). There are different recent theoretical works in support of both scenarios \cite{Raghu:10,Huo:13,Wang:13,Tsuchiizu:15}.

Recent weak-coupling renormalization group (RG) calculations \cite{Scaffidi:14,Huang:16} found the leading $p$-wave state to develop similar gap amplitudes on all three bands. In these calculations, the limit $U/W \ll 1$ where $W$ is the bandwidth (or $U/t \ll 1$ where $t$ is the primary hopping amplitude) \cite{Raghu:10a} is taken. Under the assumption that the bare onsite Coulomb interactions do not affect (see later) the superconducting solutions resultant from particle-hole density wave fluctuations, for $U/t \ra 0$ it suffices to perform a calculation up to one-loop level and the solution is considered asymptotically exact in the absence of competing particle-hole instabilities. %Justification for weak coupling in \SRO~is based on the observation of Fermi liquid behavior above $T_c$ \cite{Hussey:98}.

However, in reality Coulomb interactions in \SRO~may not be that weak. In particular, the bare interactions in multiband or multi-orbital systems may contribute either repulsive or attractive effective interactions in certain pairing channels, which may affect the predictions based on density-wave-fluctuation mechanisms. Additionally, higher order scatterings associated with the finite interactions are expected to introduce corrections to the structure of the interactions. This may be particularly important in the present multiband system, wherein multiple pairing instabilities likely coexist. To this end, we perform systematic calculations using random phase approximation (RPA) to analyze the leading pairing symmetries as a function of interaction strength, from weak to nearly intermediate interactions. The important microscopic details such as the spin-orbit coupling are fully accounted for. %This approximation allows us to generalize the weak-coupling RG calculations to larger $U$.
Our most important finding is an emerging trend: a crossover from $p$-wave pairing at extremely weak $U$ to highly anisotropic even-parity $s$- and $d$-wave pairings at relatively stronger and likely more physically relevant $U$. Our results therefore hold important implications for the pairing theories of \SRO.

%We also find that the preference between chiral and helical $p$-wave states in the extreme weak-$U$ limit is sensitive to both Hund's coupling and the inter-orbital mixing.

%\section{Model}
%\label{sec:model}
{\it Model} -- The band structure of \SRO~is described by the following three-orbital tight-binding Hamiltonian on a square lattice,
\begin{equation}
H = \sum_{{\bs k},s} \psi^\dagger_{{\bs k},s} \hat{H}_{0s}({\bs k})\psi_{{\bs k},s} \,,
\label{eq:H0}
\end{equation}
where the spinor $\psi_{{\bs k},s} = (c_{xz{\bs k},s},c_{yz{\bs k},s},c_{xy{\bs k},-s})^T$ with $c_{a{\bs k},s}$ annihilating a spin-$s$ electron on the $a$-orbital ($a=xz,yz,xy$), $s=\ua~\text{and}~\da$ denote up and down spins, and,
\begin{equation}
\hat{H}_{0s}({\bs k}) = \begin{pmatrix}
\xi_{xz,\bs k}   &  \lambda_{\bs k} - is\eta  & i\eta  \\
\lambda_{\bs k} + is\eta  & \xi_{yz,\bs k}   &  -s\eta \\
-i\eta & -s\eta & \xi_{xy,\bs k}
\end{pmatrix} \,,
\label{eq:H0a}
\end{equation}
with $\xi_{xz,\bs k} = -2t \cos k_x -2 \tilde{t} \cos k_y -\mu$, $\xi_{yz,\bs k} = -2\tilde{t} \cos k_x -2 t \cos k_y-\mu$, $\lambda_{\bs k} = -4t^{\prime\prime} \sin k_x \sin k_y $, $\xi_{xy,\bs k} = -2t^\prime(\cos k_x + \cos k_y) -4t^{\prime\prime\prime} \cos k_x \cos k_y -\mu_1$. Here $\lambda_{\bs k}$ is the inter-orbital hybridization between the two quasi-1D $xz$- and $yz$-orbitals; and $\eta$ is the strength of spin-orbit coupling (SOC) which is found to be sizable \cite{Haverkort:08,Veenstra:14,Fatuzzo:15}. Note that because SOC mixes different spins on the $xy$- and the other two orbitals, the spins are not good quantum numbers. However, since the Kramers degeneracy on each band is preserved, it is convenient to adopt a pseudospin notation where the electrons on the Bloch bands are denoted pseudospin-up and down fermions. We fix the band parameters: $(t,\tilde{t},t^\prime,t^{\prime\prime\prime},\mu,\mu_1) = (1,0.1,0.8,0.3,1,1.1)t$ which are known to capture the overall band structure and Fermi surface geometry of \SRO~\cite{Damascelli:00,Bergemann:03,Bergemann:00}. For now we leave undetermined the magnitude of the orbital mixing $t^{\prime\prime}$ and $\eta$. Their values will be suitably tuned to analyze the influence of the associated microscopic details.

We consider the onsite Coulomb interactions between the Ru $t_{2g}$ orbitals as the following,
\begin{eqnarray}
H_\text{int}&=& \sum_{i,a,s\neq s^\prime}\frac{U}{2} n_{ias} n_{ias^\prime} + \sum_{i,a\neq b,s,s^\prime} \frac{U^\prime}{2}n_{ias}n_{ibs^\prime} \nonumber \\
&+& \sum_{i,a\neq b,s,s^\prime} \frac{J}{2} c^\dagger_{ias}c^\dagger_{ibs^\prime}c_{ias^\prime}c_{ibs} \nonumber \\
&+& \sum_{i,a\neq b,s\neq s^\prime} \frac{J^\prime}{2}c^\dagger_{ias}c^\dagger_{ias^\prime}c_{ibs^\prime}c_{ibs} \,,
\label{eq:interactions}
\end{eqnarray}
where $i$ is the site index, $a,b=xz,yz,xy$, $n_{ias} \equiv c^\dagger_{ias}c_{ias}$. Throughout this study we assume $U^\prime=U-2J$ and $J^\prime=J$ where $J$ is the Hund's coupling.

To study the superconducting instabilities, we obtain effective pairing vertices using systematic RPA calculations formulated in the pseudospin language \cite{seeSupp}, which differ from the previous RPA and perturbative expansion studies \cite{Takimoto:00,Nomura:00,Nomura:02}. Note that as the atomic SOC does not break inversion symmetry, notions of odd- and even-parity pairings remain valid and are in one to one correspondence with {\it pseudospin-triplet} and {\it pseudospin-singlet} pairings. Accordingly, the obtained gap functions acquire the standard forms $\hat{\Delta}^t_{\bs k}\sim i (\bs \sigma \cdot \bs d_{\bs k})\sigma_y$ and $\hat{\Delta}^s_{\bs k}\sim i \Delta_{\bs k} \sigma_y$, which are expressed in the pseudospin basis. The forms of ${\bs d}_{\bs k} $ and $\Delta_k$ fall in the irreducible representations of the $D_{4h}$ crystalline symmetry group, and the orientation of the $\bs d$-vector ($\bs d_{\bs k}$) represents the pseudospin configuration of the odd-parity pairing. The simple chiral $p$-wave is given by $(k_x \pm ik_y) \hat{z}$, while the helical $p$-wave states are marked by in-plane ${\bs d}$ orientations: $k_x \hat{x} \pm k_y \hat{y}$ and $k_y \hat{x} \pm k_x \hat{y}$.

\begin{figure}
\includegraphics[width=8.5cm]{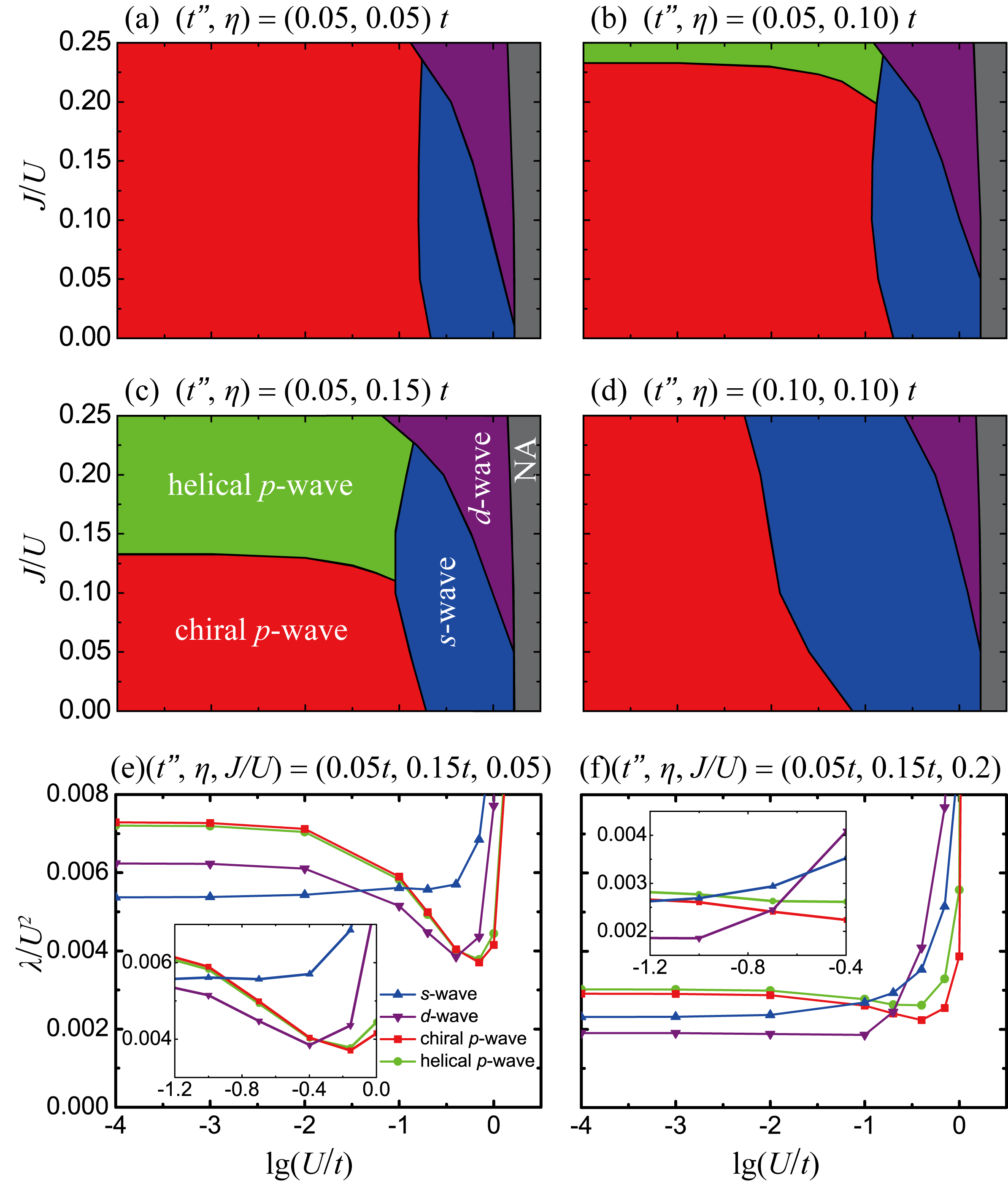} %{PhaseDiagNew3c.png}
\caption{(color online) Phase diagrams as a function of the interaction parameters $J/U$ and $\lg(U/t)$ for $(t^{\prime\prime},\eta)=(0.05,0.05)t$ [a], $(0.05,0.10)t$ [b], $(0.05,0.15)t$ [c] and $(0.10,0.10)t$ [d]. Only the single most leading channel at each point is indicated (see text in [c] for labels): chiral $p$-wave (red), helical $p$-wave (green), $s$-wave (blue) and $d_{x^2-y^2}$-wave (purple), and `NA' (grey) corresponds to the regime where the RPA susceptibility diverges and where our method breaks down. The phase boundaries are rough estimates. They are smoothed lines connecting the approximate midpoints which separate our data points (see \onlinecite{seeSupp}) associated with different phases. (e-f), evolution of the eigenvalues of the gap equation in different channels as a function of the interaction parameter $\text{lg}(U/t)$. Note the log-scale in $x$-axis.}
\label{fig:PhaseDiag}
\end{figure}

To deduce the general behavior, we perform calculations using orbital mixing parameters in the range $t^{\prime\prime},\eta \in (0,0.15)t$. Figure \ref{fig:PhaseDiag} depicts the phase diagrams in terms of the dimensionless interaction parameters $J/U$ and $\text{lg}(U/t)$ for four sets of $(t^{\prime\prime},\eta)$, which are roughly representative of our overall observation. In the following we separately discuss the two limiting cases: extreme weak-$U$ [i.e. $\text{lg}(U/t)<-1$ or $U<0.1t$] and finite-$U$ [i.e. $\text{lg}(U/t)>-1$].

\begin{figure}
\includegraphics[width=8.7cm]{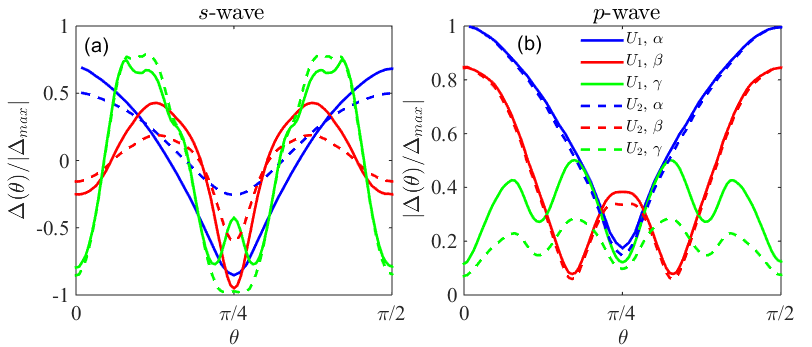}
\caption{(color online) The $s$- and $p$-wave gaps as a function of the angle with respect to the $k_y$-axis, at two different interaction parameters $U=U_1=0.01t$ (solid) and $U=U_2=0.2t$ (dashed). The calculations are done with $(t^{\prime\prime},\eta)=(0.05,0.15)t$ and $J/U=0.1$. Note that $s$-wave is the third and first leading state at $U_1$ and $U_2$, respectively; while $p$-wave is the first and second.}
\label{fig:GapMagnitude}
\end{figure}

{\it Weak-$U$ limit} -- In the extreme weak coupling limit, the $p$-wave gap functions obtained in our calculations are in excellent agreement with the previous study \cite{Scaffidi:14} [see Fig \ref{fig:GapMagnitude} (b)]. In the presence of sizable SOC, the three bands are more prompt to develop comparable gaps. The SOC also induces anisotropic spin correlations \cite{Ng:00,Eremin:02,Cobo:16} responsible for the splitting between the chiral and helical channels. Naturally, the balance between the two is sensitive to SOC, which acts in conjunction with other microscopic details in the band structure and the bare interactions, such as $t^{\prime\prime}$ and $J/U$. Crudely speaking, chiral $p$-wave state wins over helical $p$-wave at smaller $J/U$, and stronger SOC tips the balance towards helical $p$-wave [Fig \ref{fig:PhaseDiag} (a)-(c)]. The former is consistent with the previous study which found leading chiral and helical pairings for small and larger $J/U$, respectively. In addition, at larger $t^{\prime\prime}$, the chiral state develops more favorably than the helical pairing, and more $p$-wave phase space becomes overtaken by even-parity pairing [see Fig \ref{fig:PhaseDiag} (b) and (d)]. The sensitivity to the microscopics was also noted in a recent work \cite{Hsu:17}.

As a crucial remark, at the RPA level treated here the spin fluctuations {\it on their own} in fact favor even-parity $s$- or $d$-wave pairing, instead of $p$-wave. In these extreme weak-$U$ calculations, $p$-wave surpasses the others because the even-parity channels are suppressed by nonvanishing repulsive components at the bare-$U$ level. We have verified this via explicit calculations where only the bare-$U$ contribution is included and where it is purposely taken out. Hence caution is needed when interpreting the extreme weak-$U$ results.

%As a crucial remark, the peculiar fragility of the leading pairing symmetry against the microscopics raises a critical question regarding the reliability of the extreme weak-$U$ (and one-loop) calculations.

\begin{figure}
\subfigure{ \includegraphics[width=8.7cm]{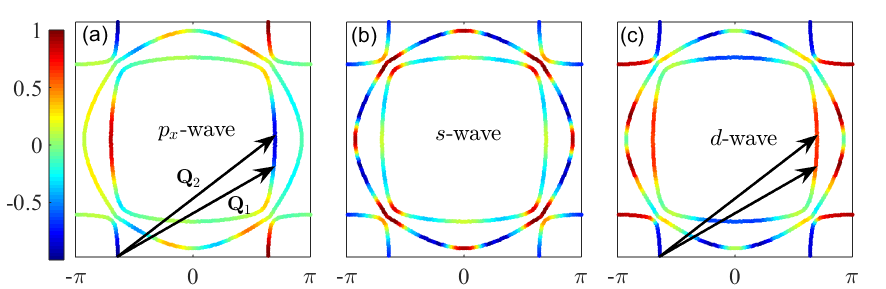} }
\subfigure{ \includegraphics[width=8.7cm]{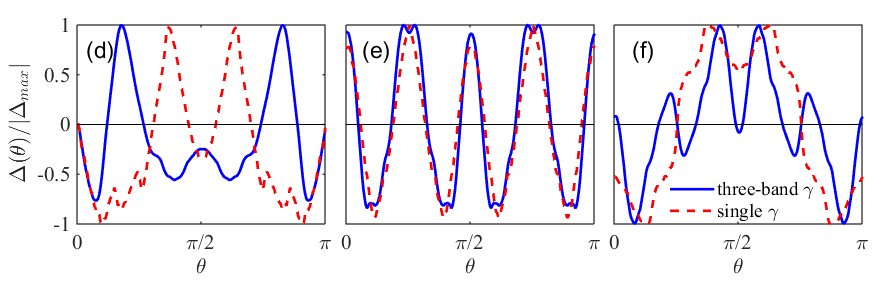} }
\caption{(color online) Upper panel: leading RPA gap functions in three regions of interaction strength: (a) $U=10^{-4}t$, (b) $U=0.1t$ and (c) $U=t$, wherein the three-band model give most leading chiral $p$-wave (shown is the $p_x$-component), $s$- and $d_{x^2-y^2}$-wave pairings, respectively. Lower panel: comparison of $\gamma$-band gap functions obtained in the three-band model (solid) and an effective one-band model (dashed), under the same interaction strength in correspondence with those in the upper panels. In the plots shown, the three-band model assumes interband mixing $(t^{\prime\prime},\eta)=(0.05,0.15)t$ and the interaction parameter $J/U$ is held fixed at 0.1. The one-band model uses a band structure and filling fraction that resemble those of the $\gamma$-band. Its leading channel is always $d$-wave at the chosen filling level, $\mu_1=1.1t$ (see \onlinecite{seeSupp}), with subdominant $p$- and $s$-wave pairings. The wavevectors ${\bs Q_1}$ and ${\bs Q_2}$ shown in (a) and (c) are approximately equal to those mentioned in the text, up to a reciprocal lattice wavevector.}
\label{fig:onebandGamma}
\end{figure}

%\begin{figure}
%\subfigure{ \includegraphics[width=2.5cm]{Gaptpp0p05eta0p1Px.png} }
%\subfigure{ \includegraphics[width=2.5cm]{Gaptpp0p05eta0p1S.png} }
%\subfigure{ \includegraphics[width=2.9cm]{Gaptpp0p05eta0p1D1.png} }
%\caption{(color online) Most leading gap functions }
%\label{fig:Gaps}
%\end{figure}

{\it Intermediate-$U$} -- Most important of all, a robust emerging trend, irrespective of the details of the orbital mixing, is the crossover at relatively weak interaction strength from $p$-wave to $s$- and $d$-wave pairings which are in the $A_{1g}$ and $B_{1g}$ representations, respectively. In addition, the $d$-wave pairing is increasingly favored at larger $J/U$  (Fig \ref{fig:PhaseDiag}), and the $s$-wave channel invariably exhibits strong gap anisotropy with multiple accidental nodes across the Fermi surfaces [e.g. Fig \ref{fig:GapMagnitude} (a) and Fig \ref{fig:onebandGamma} (b))]. These constitute the central message of the present study. Some of the representative gap functions are depicted in Fig \ref{fig:onebandGamma} (a-c).

Typically at the order of $U =U_0\sim 0.1t$ [Fig \ref{fig:PhaseDiag} (e-f)], the spin-fluctuation-mediated effective interactions begin to supersede the bare-$U$ repulsion in several even-parity channels. Note that although at this order the bare-$U$ interactions are still relatively strong compared with $\mathcal{O}(U^2)$, their components in the respective eigen channels can be much weaker. Across $U_0$, the obtained gap structure exhibits quantitative variations in, e.g. the detailed shape and the relative amplitudes of the gaps across the three bands, as is illustrated in Fig \ref{fig:GapMagnitude}.

Since the bare-$U$ interactions have no effect on the $p$-wave pairing, variations in this channel must result exclusively from the higher order corrections in RPA. The inaction of bare-$U$ permits a stronger statement, that the $p$-wave pairing must be driven by spin fluctuations alone. Traces of this mechanism \cite{Tsuchiizu:15,Scaffidi:15} can be found in, e.g. the approximate relation $\text{sgn}[\Delta_{\bs k}] = \text{sgn}[\Delta_{\bs k+\bs Q}]$ in the near-nested portions of the quasi-1D Fermi surfaces, as indicated by the arrows in Fig \ref{fig:onebandGamma} (a). A similar degree of RPA correction is expected for even-parity pairings. However, since the bare-$U$ repulsion and the spin-fluctuation-mediated interactions `interfere' nontrivially in these channels, it was not until at somewhat larger $U$ where the latter dominates. For example, as shown in Fig \ref{fig:onebandGamma} (c), for a model with $U=t$ which develops sizable pairing gaps on the quasi-1D bands, the nested regions satisfy $\text{sgn}[\Delta_{\bs k}] = -\text{sgn}[\Delta_{\bs k+\bs Q}]$ characteristic of an even-parity pairing favored by the momentum-${\bs Q}$ spin fluctuations.

However, care must be taken when attributing the superconductivity to certain spin fluctuation modes, as the interactions in this multi-orbital system are inherently multiband in nature (except in the strong-$U$ limit where a single mode prevails) \cite{Huang:16}. The multiband effects may operate differently in different channels, which is best illustrated by comparing the $\gamma$-band pairing in single- and multiband models. Its van Hove singularity induces spin fluctuations not only at small momenta, but also at large wavevectors surrounding $(\pi,\pi)$ \cite{Liu:17}. Depending on the interaction strength and the exact filling fraction, these fluctuations may support $s$-, $p$-, or $d$-wave channels within RPA \cite{seeSupp}. The gap functions are plotted along with the multiband results in Fig \ref{fig:onebandGamma} (d-f). Obviously, distinctions appear not only around the locus of maximal band-mixing, but also away from them, despite very similar $\gamma$-band spin fluctuations in the two models \cite{seeSupp}. Especially, the contrast is much more appreciable in the $p$- and $d$-wave channels, suggesting stronger multiband effects at play in the two.

%Notwithstanding, the distribution of the $s$- and $d$-wave gaps on the three bands is not generic.

It shall be stressed that the multiband character persists, irrespective of the distribution of the gaps on the three bands which, by contrast, is not generic. Depending on the parametrization of the orbital mixing, the pairing can reside primarily on one set of bands or be of similar magnitude on all three. On the one hand, stronger SOC in general leads to more comparable gap amplitudes; on the other hand, since the momentum-dependent $xz/yz$-orbital hybridization $\lambda_{\bs k}$ destroys the $\alpha/\beta$-band nesting conditions more dramatically, with larger $t^{\prime\prime}$ the pairing on these two bands are increasingly suppressed compared to that on $\gamma$.

%Finally, at the strongest $U$ where RPA is still valid, higher order scatterings introduce substantial corrections to the momentum-space structure of the interactions. In principle, the strongest corrections appear at transfer momenta coincident with the peak positions of the spin fluctuations. Typically either of the two sets of bands dominates pairing in this limit, which is indicative of the influence of either of the most prominent Fermi surface features mentioned previously. In this case it shall be justifiable to use models which treat the 1D and 2D bands separately.

{\it Other symmetries} -- Planar chiral $f$-wave pairings in product representations, taking the forms of $(k_x^2 -k_y^2)(k_x+ik_y)\hat{z}$ and $k_xk_y(k_x+ik_y)\hat{z}$ \cite{Hasegawa:00,Graf:00}, were proposed to explain the reported low temperature thermodynamic and transport measurements \cite{Ishida:97,NishiZaki:00,Hassinger:16}. However, we did not find these pairings among the most leading channels obtained in our calculations.

%To understand this more qualitatively, we turn to the effective multiband interactions, defined as ...,

%Verify once again that the $p$-wave pairing on the quasi-1D bands, if any, is driven by the relatively stronger interband interaction between them; and that their respective intraband interactions disfavor this pairing channel. Meanwhile, the Cooper instability on the  $\gamma$-band is ensured by a noticeable intraband attraction, supplemented by comparatively weaker interactions with the other bands.

% In different symmetry channels, the configuration of the effective multiband interaction are markedly different. The multiband interaction matrix indicates that, while the quasi-nesting between the quasi-1D bands results in strong interband interactions between the two bands in both triplet and singlet channels, the singlet pairing on the 1D bands is disfavored by similarly strong intraband repulsions. By contrast, the triplet channel sees much weaker intraband repulsions on the two bands compared to the interband interaction between them.

%5) Near-degeneracy between $d_{x^2-y^2}$ and $d_{xy}$ channels? Does this occur in a broad parameter range, in the presence of finite $U$? \\

%8) Leggett mode assisted Josephson tunneling. \\

{\it Implications for RG studies} -- In the $U/t \ra 0$ limit within the regime of validity of the two-stage weak-coupling RG developed in Ref. \onlinecite{Raghu:10}, one first integrates out the states with energies higher than an artificial cutoff $\Omega_0 \ll U^2/t$. This generates a low-energy effective action, from which one computes the RG flow of the individual Cooper channels. Since $\Omega_0 \ra 0$, we expect the existing weak-coupling RG study as formulated in Ref \onlinecite{Scaffidi:14} to yield the same results as our RPA, as it did. In the intermediate range of $U$ still far from particle-hole instabilities, e.g. $0.1<U/t < \mathcal{O}(1)$ in the present study \cite{footnote}, the system can still be viewed as weakly coupled. However, the perturbative treatment of the high-energy modes must now be carried out more accurately. For example, Zanchi and Schulz \cite{Zanchi:96} implemented a standard one-loop RG calculation for the high-energy states down to a physical cutoff, below which the flow of the particle-hole loops disappear exponentially. Given the nontrivial interplay of the bare-$U$ and higher-order interactions found in RPA, a corresponding study within RG is naturally of considerable interest and will be pursued separately.

The functional RG (fRG) approaches \cite{Wang:13,Tsuchiizu:15} by contrast use sizable $U$. In addition, they take into account the interplay between the pairing and the particle-hole channels in the RG flow--an ingredient missing in weak-coupling RG. In these studies, the growth of the odd-parity pairing channel is found to be faster than that of the even-parity ones as the temperature or cutoff energy decreases. However, although both yield leading $p$-wave, they disagree on which set of bands drives the pairing. Inferring from our results, this could potentially originate from the different microscopic details adopted in the two studies. It therefore seems important to attend to the microscopics and scan over a broader parameter range in order to obtain a complete phase diagram.

{\it Concluding remarks} -- The leading pairing obtained in our RPA appears in the $p$-wave channel in the extreme weak-coupling regime, and in even-parity $s$- or $d$-wave channels in a broad range of intermediate interactions -- in contrast with the widely-held assumption for \SRO. We do note that multiple signatures inconsistent with chiral $p$-wave may go very well with even-parity pairing. For example, taken at face value, these pseudospin singlet states are in line with the suppression of the in-plane $H_{c2}$ and the character of first-order superconducting transition at low temperatures \cite{Yonezawa:13,Yonezawa:14,Kittaka:14,Deguchi:02,Amano:15,Nakai:15}, although these behavior may also be explained by invoking an orbital polarization mechanism due to finite SOC \cite{Ramires:16}. Moreover, the obtained gap functions in these channels exhibit multiple accidental ($s$-wave) or symmetry-imposed ($d$-wave) vertical line nodes, which may explain the residual density of states \cite{NishiZaki:00,Hassinger:16,Ishida:97}. In addition, on a tetragonal lattice as in \SRO, planar even-parity pairing belongs to 1D irreducible representation and typically does not break time-reversal invariance; hence the superconductor is not expected to generate chiral current at the edges, nor split transitions in the presence of $C_4$ symmetry breaking perturbations. Although it is still premature to draw a firm conclusion of alternative pairing symmetry in this material due to the obvious reason that even-parity pairing cannot be straightforwardly reconciled with the many observations in support of $p$-wave pairing, our results call for new perspectives in studying the superconductivity in \SRO. As a side remark, our work also shows how the pairing in a multiband superconductor can be strongly influenced by the multiband interactions. 

{\it Acknowledgements} -- We would like to acknowledge helpful communications with John Berlinsky, Ilya Eremin, Catherine Kallin, Kazushige Machida, Thomas Scaffidi, Manfred Sigrist, Steve Simon, Yukio Tanaka and Fuchun Zhang. This work is supported by NSFC under Grant No. 11604013 (LDZ) and Grant Nos. 11674025, 11334012, 11274041 (FY), Beijing Natural Science Foundation under Grant No. 1174019 (LDZ), and the C. N. Yang Junior Fellowship at Tsinghua University (WH).

\newpage

\begin{center}
\textbf{\large Supplemental Material: Superconducting pairing in Sr$_2$RuO$_4$ from weak to intermediate coupling}
\end{center}

\renewcommand{\theequation}{S\arabic{equation}}
\setcounter{equation}{0}
\renewcommand{\thefigure}{S\arabic{figure}}
\setcounter{figure}{0}
\renewcommand{\thetable}{S\arabic{table}}
\setcounter{table}{0}

\section{I. Methods}
\subsection{A. Coulomb interactions}
The interaction Hamiltonian adopted in our calculations is
\begin{align}\label{Hint1}
H_{int}=&\frac{U}{2}\sum_{i,a,s\neq s'}
n_{ias}n_{ias'}
+\frac{U'}{2}\sum_{i,a\neq b,s,s'}
n_{ias}n_{ibs'}                      \nonumber\\
&+\frac{J}{2}\sum_{i,a\neq b,s,s'}
c^{\dagger}_{ias}
c^{\dagger}_{ibs'}
c_{ias'}c_{ibs}                      \nonumber\\
&+\frac{J}{2}\sum_{i,a\neq b,s\neq s'}
c^{\dagger}_{ias}
c^{\dagger}_{ias'}
c_{ibs'}c_{ibs}
\end{align}
where $i$ is the site index, $a,b=xz,yz,xy$ are the orbital indices, $s,s'=\uparrow,\downarrow$ are the real-spin indices, and $U'=U-2J$. To simplify the forms of the formula, we define $U^{as_abs_b}_{cs_cds_d}$ having the following nonzero elements:
\begin{align}\label{U1}
U^{as bs}
_{c\bar{s}d\bar{s}}
=\left\{
\begin{array}{ll}
U,  & a=b=c=d; \\
U', & a=b\neq c=d; \\
J, & a=c\neq b=d; \\
J,  & a=d\neq c=b,
\end{array}
\right.
\end{align}
\begin{align}\label{U2}
U^{as b\bar{s}}
_{c\bar{s}ds}
=\left\{
\begin{array}{ll}
-U,  & a=b=c=d; \\
-J, & a=b\neq c=d; \\
-J, & a=c\neq b=d; \\
-U',  & a=d\neq c=b,
\end{array}
\right.
\end{align}
\begin{align}\label{U3}
U^{as bs}
_{cs ds}
=\left\{
\begin{array}{ll}
U'-J,  & a=b\neq c=d;  \\
-U'+J, & a=d\neq c=b,  \\
\end{array}
\right.
\end{align}
where $\bar{s}\equiv-s$, and thus satisfying
\begin{align}\label{Usym}
U^{as_abs_b}_{cs_cds_d}=U^{cs_cds_d}_{as_abs_b}
=-U^{as_ads_d}_{cs_c bs_b}=-U^{cs_c bs_b}_{as_ads_d}.
\end{align}
Then, we can write the interaction Hamiltonian (\ref{Hint1}) as
\begin{align}\label{Hint2}
H_{int}=\frac{1}{4}\sum_{iabcds_as_bs_cs_d}
U^{as_abs_b}_{cs_cds_d}
c^{\dagger}_{ias_a}c_{ibs_b}
c^{\dagger}_{ics_c}c_{ids_d}.
\end{align}
Since spin-orbit coupling mixes opposite spins from the quasi-1D and quasi-2D orbitals, it is no longer appropriate to speak about Cooper pairing in real-spin space. However, given the preserved Kramers degeneracy on the bands, it is convenient to define the electrons on the Bloch bands as pseudospin- up and down fermions. It is then perfectly valid to use notions of pseudospin pairings. We thus rewrite the interaction Hamiltonian (\ref{Hint2}) in the pseudo-spin basis as follows:
\begin{align}\label{Hint3}
H_{int}=\frac{1}{4}\sum_{iabcd\sigma_a\sigma_b\sigma_c\sigma_d}
U^{a\sigma_ab\sigma_b}_{c\sigma_cd\sigma_d}
c^{\dagger}_{ia\sigma_a}c_{ib\sigma_b}
c^{\dagger}_{ic\sigma_c}c_{id\sigma_d}.
\end{align}
Here the pseudo-spin index $\sigma_a=s_a$ for $a=xz,yz$; and $\sigma_a=\bar{s}_a$ for $a=xy$, which also applies to the pseudospin indices $\sigma_b$, $\sigma_c$, and $\sigma_d$. As no more reference to real spins is needed hereafter, from now on we use `$\ua$' and `$\da$' symbols to exclusively designate pseudospin up and down, and drop the prefix `pseudo' to simplify the notation, unless otherwise specify.

\subsection{B. Susceptibility}
To perform the RPA calculation for Sr$_2$RuO$_4$, we need to extend the standard RPA approach \cite{RPA1,RPA2,RPA3,Kuroki,Scalapino1,Scalapino2,Liu2013,Wu2014,Ma2014,Zhang2015} to a more general form. To this end, we first define the generalized susceptibility of the system as
\begin{align}
\chi^{a\sigma_ab\sigma_b}
_{c\sigma_cd\sigma_d}(\bm{q},\tau)
\equiv\frac{1}{N}&\sum_{\bm{k}_1\bm{k}_2}
\left\langle T_{\tau}c^{\dagger}_{a\sigma_a}(\bm{k_1},\tau)
c_{b\sigma_b}(\bm{k_1}+\bm{q},\tau)\right.                      \nonumber\\
&\left.\times c^{\dagger}_{c\sigma_c}(\bm{k_2}+\bm{q},0)
c_{d\sigma_d}(\bm{k_2},0)\right\rangle
\end{align}
where $T_{\tau}$ denotes the time-ordered product, and $\langle\cdots\rangle$ denotes the thermal average of the system. When the interaction is turned off ($U=U'=J=0$), the above generalized susceptibility reduces to the bare susceptibility $\chi^{(0)a\sigma_ab\sigma_b}
_{~~~c\sigma_cd\sigma_d}(\bm{q},\tau)$, which is nonzero only if $\sigma_d=\sigma_a$ and $\sigma_c=\sigma_b$. Fourier transformed to the imaginary frequency space, the nonzero elements of the bare susceptibility can be expressed by the following explicit form:
\begin{align}
\chi&^{(0)a\sigma_ab\sigma_b}
_{~~~c\sigma_bd\sigma_a}(\bm{q},iw_n)
=\frac{1}{N}\sum_{\bm{k}\alpha\beta}
\xi^{\alpha}_{d\sigma_a}(\bm{k})
\xi^{\alpha*}_{a\sigma_a}(\bm{k})                      \nonumber\\
&\times \xi^{\beta}_{b\sigma_b}(\bm{k}+\bm{q})
\xi^{\beta*}_{c\sigma_b}(\bm{k}+\bm{q})
\frac{n_F(\varepsilon^{\beta}_{\bm{k}+\bm{q}})
-n_F(\varepsilon^{\alpha}_{\bm{k}})}
{iw_n+\varepsilon^{\alpha}_{\bm{k}}
-\varepsilon^{\beta}_{\bm{k}+\bm{q}}},
\label{eq:chi0}
\end{align}
where $\alpha,\beta$ are band indices, $\varepsilon^{\alpha}_{\bm{k}}$ is the $\alpha$-band dispersion, $\xi^{\alpha}_{l}\left(\bm{k}\right)$ is the matrix element of the obital-to-band transformation, $n_F$ is the Fermi-Dirac distribution function, and $w_n = 2n\pi/T$ with $n$ integer is the Matsubara frequency. When the interaction is turned on, the generalized susceptibility at the RPA level reads
\begin{align}\label{RPA}
\chi(\bm{q},iw_n)=\left[1+\chi^{(0)}(\bm{q},iw_n)(U)\right]^{-1}
\chi^{(0)}(\bm{q},iw_n).
\end{align}
where $\chi(\bm{q},iw_n)$, $\chi^{(0)}(\bm{q},iw_n)$, and $(U)$ operate as $36\times36$ matrices with $(U)^{a\sigma_a b\sigma_b}_{c\sigma_cd\sigma_d}
\equiv U^{a\sigma_ab\sigma_b}_{c\sigma_cd\sigma_d}$. The plus sign in the square braket, though seemingly different than the usual conventions used elsewhere, is in fact due to the particular structure of the interaction matrix chosen here. Note this approach only works for interactions at which all the eigenvalues of the denominator
matrix $\left[1+\chi^{(0)}(\bm{q},iw_n)(U)\right]$ are positive. When $U$ reaches a $J/U$-dependent critical value $U_c$ at which the lowest eigenvalue of the denominator matrix becomes zero, the generalized susceptibility (\ref{RPA}) diverges, thus the RPA treatment will breakdown. Throughout this work, we use $T=0.001t$ and a $400\times 400$ $\bs k$-mesh to evaluate the susceptibility.

%Note the distinction between the pseudo-spin susceptibilities evaluated here and those measured in neutron scattering or calculated in real-spin representation in, e.g. Refs \onlinecite{Eremin:02,Cobo:16}. Although the former does exhibit certain features of the real-spin susceptibility, ... 

\subsection{C. Effective interaction}
At small interaction strength $U<U_c$, the system exhibits short range spin and charge fluctuations. The exchange of these fluctuations leads to the pairing interaction responsible for superconductivity. Taking the static limit by sending $w_n=0$, for the present system the pairing interaction vertex at the RPA level reads
\begin{align}\label{gamma}
\Gamma^{a\sigma_ab\sigma_b}_{c\sigma_cd\sigma_d}
(\bm{k},\bm{k}')
=&(U)^{a\sigma_ad\sigma_d}_{b\sigma_bc\sigma_c}
-\left[(U)\chi(\bm{k}-\bm{k}')(U)\right]
^{a\sigma_ad\sigma_d}_{b\sigma_bc\sigma_c}                      \nonumber\\
&+\left[(U)\chi(\bm{k}+\bm{k}')(U)\right]
^{a\sigma_ac\sigma_c}_{b\sigma_bd\sigma_d}.
\end{align}
Projecting this interaction onto the band basis, and considering only the intra-band pairing one arrives at the following effective interaction on the Fermi level:
\begin{align}
V_\text{eff}=&\frac{1}{N}\sum_{\bm{k}\bm{k}'\alpha\beta
\sigma_a\sigma_b\sigma_c\sigma_d}
V^{\alpha\beta}_{\sigma_a\sigma_b\sigma_c\sigma_d}(\bm{k},\bm{k}')            \nonumber\\
&\times c^{\dagger}_{\alpha\sigma_a}(\bm{k})
c^{\dagger}_{\alpha\sigma_b}(-\bm{k})
c_{\beta\sigma_c}(-\bm{k}')c_{\beta\sigma_d}(\bm{k}'),
\end{align}
where
\begin{align}
V^{\alpha\beta}_{\sigma_a\sigma_b\sigma_c\sigma_d}&(\bm{k},\bm{k}')
=\frac{1}{4}\text{Re}\sum_{abcd}\Gamma^{a\sigma_ab\sigma_b}
_{c\sigma_cd\sigma_d}(\bm{k},\bm{k}')                      \nonumber\\
&\times\xi^{\alpha*}_{a\sigma_a}(\bm{k})
\xi^{\alpha*}_{b\sigma_b}(-\bm{k})
\xi^{\beta}_{c\sigma_c}(-\bm{k}')
\xi^{\beta}_{d\sigma_d}(\bm{k}').
\end{align}

\subsection{D. Linearized gap equation}
For convenience we take pseudospin quantization in the $z$-direction. The superconducting pairing of the system can then be divided into two categories: the opposite-spin pairing (OSP) and the equal-spin pairing (ESP). The solutions in the former can be further classified into the even- and odd-parity channels, while solutions in the latter category belongs exclusively to odd-parity states. In the odd-parity channels, an OSP with odd-parity is a state with ${\bs d}$-vector along $z$, and an ESP state is characterized by in-plane ${\bs d}$.  For the OSP states, we define the symmetrized effective interaction \begin{align}
V^{\alpha\beta}_\text{OSP}(\bm{k},\bm{k}')
=&V^{\alpha\beta}_{\uparrow\downarrow\downarrow\uparrow}(\bm{k},\bm{k}')
+V^{\alpha\beta}_{\downarrow\uparrow\uparrow\downarrow}(-\bm{k},-\bm{k}')   \nonumber\\
&-V^{\alpha\beta}_{\uparrow\downarrow\uparrow\downarrow}(\bm{k},-\bm{k}')
-V^{\alpha\beta}_{\downarrow\uparrow\downarrow\uparrow}(-\bm{k},\bm{k}'),
\end{align}
and the gap function
\begin{align}
\Delta^{\alpha}_{\bm{k}}=&\frac{1}{N}
\sum_{\bm{k}'\beta}V^{\alpha\beta}(\bm{k},\bm{k}')
\Big\langle c_{\beta\downarrow}(-\bm{k}')
c_{\beta\uparrow}(\bm{k}')\Big\rangle.
\end{align}
They satisfy the following linearized gap equation near $T_c$:
\begin{align}
-\frac{1}{(2\pi)^2}\sum_{\beta}\oint_{FS}d\bm{k}'_{\parallel}
\frac{V^{\alpha\beta}_\text{OSP}(\bm{k},\bm{k}')}
{v^{\beta}_F(\bm{k}')}\Delta^{\beta}_{\bm{k}'}
=\lambda\Delta^{\alpha}_{\bm{k}},
\end{align}
where $\lambda$ is the pairing eigenvalue. For the ESP states, we define the symmetrized effective interaction
\begin{align}\label{Delta1}
V^{\alpha\beta}_\text{ESP}(\bm{k},\bm{k}')=&\left(
\begin{array}{cc}
V^{\alpha\beta}_{\uparrow\uparrow\uparrow\uparrow}(\bm{k},\bm{k}')
& V^{\alpha\beta}_{\uparrow\uparrow\downarrow\downarrow}(\bm{k},\bm{k}') \\
V^{\alpha\beta}_{\downarrow\downarrow\uparrow\uparrow}(\bm{k},\bm{k}')
& V^{\alpha\beta}_{\downarrow\downarrow\downarrow\downarrow}(\bm{k},\bm{k}') \\
\end{array}
\right)               \nonumber\\
&-\left(
\begin{array}{cc}
V^{\alpha\beta}_{\uparrow\uparrow\uparrow\uparrow}(\bm{k},-\bm{k}')
& V^{\alpha\beta}_{\uparrow\uparrow\downarrow\downarrow}(\bm{k},-\bm{k}') \\
V^{\alpha\beta}_{\downarrow\downarrow\uparrow\uparrow}(\bm{k},-\bm{k}')
& V^{\alpha\beta}_{\downarrow\downarrow\downarrow\downarrow}(\bm{k},-\bm{k}') \\
\end{array}
\right),
\end{align}
and the gap function
\begin{align}
\Delta^{\alpha}_{\bm{k}\sigma\sigma}=\frac{1}{N}
&\sum_{\bm{k}'\beta}\left[V^{\alpha\beta}
_{\sigma\sigma\sigma\sigma}(\bm{k},\bm{k}')
\Big\langle c_{\beta\sigma}(-\bm{k}')
c_{\beta\sigma}(\bm{k}')\Big\rangle\right.                         \nonumber\\
&~~~~\left.+V^{\alpha\beta}
_{\sigma\sigma\bar{\sigma}\bar{\sigma}}(\bm{k},\bm{k}')
\Big\langle c_{\beta\bar{\sigma}}(-\bm{k}')
c_{\beta\bar{\sigma}}(\bm{k}')\Big\rangle\right].
\end{align}
They satisfy the following linearized gap equation near $T_c$:
\begin{align}
-\frac{1}{(2\pi)^2}\sum_{\beta}\oint_{FS}d\bm{k}'_{\parallel}
\frac{V^{\alpha\beta}_\text{ESP}(\bm{k},\bm{k}')}{v^{\beta}_F(\bm{k}')}\left(
\begin{array}{c}
\Delta^{\beta}_{\bm{k}'\uparrow\uparrow} \\
\Delta^{\beta}_{\bm{k}'\downarrow\downarrow} \\
\end{array}
\right)
=\lambda\left(
\begin{array}{c}
\Delta^{\alpha}_{\bm{k}\uparrow\uparrow} \\
\Delta^{\alpha}_{\bm{k}\downarrow\downarrow} \\
\end{array}
\right).
\end{align}
The critical temperature $T_c$ itself is determined by the overall largest pairing eigenvalue $\lambda$ between the above solutions through $T_c\propto \Lambda e^{-1/\lambda}$ where $\Lambda$ is a cutoff energy characteristic of the width of the energy window within which the interactions distribute.

Note that in the calculations we typically choose around 1000 equally-spaced $\bs k$ points on the Fermi surface to ensure good convergence. The phase diagrams in the main text are drawn based on the data points using the following parametrization of interactions: $U=[10^{-4},10^{-3},10^{-2},10^{-1},0.2,0.4,0.7,1,1.5,2]t$, $J/U=[0.01,0.03,0.05,0.07,0.1,0.15,0.2,0.25]$.

\section{II. single $\gamma$-band model}
\begin{figure}
\includegraphics[width=6cm]{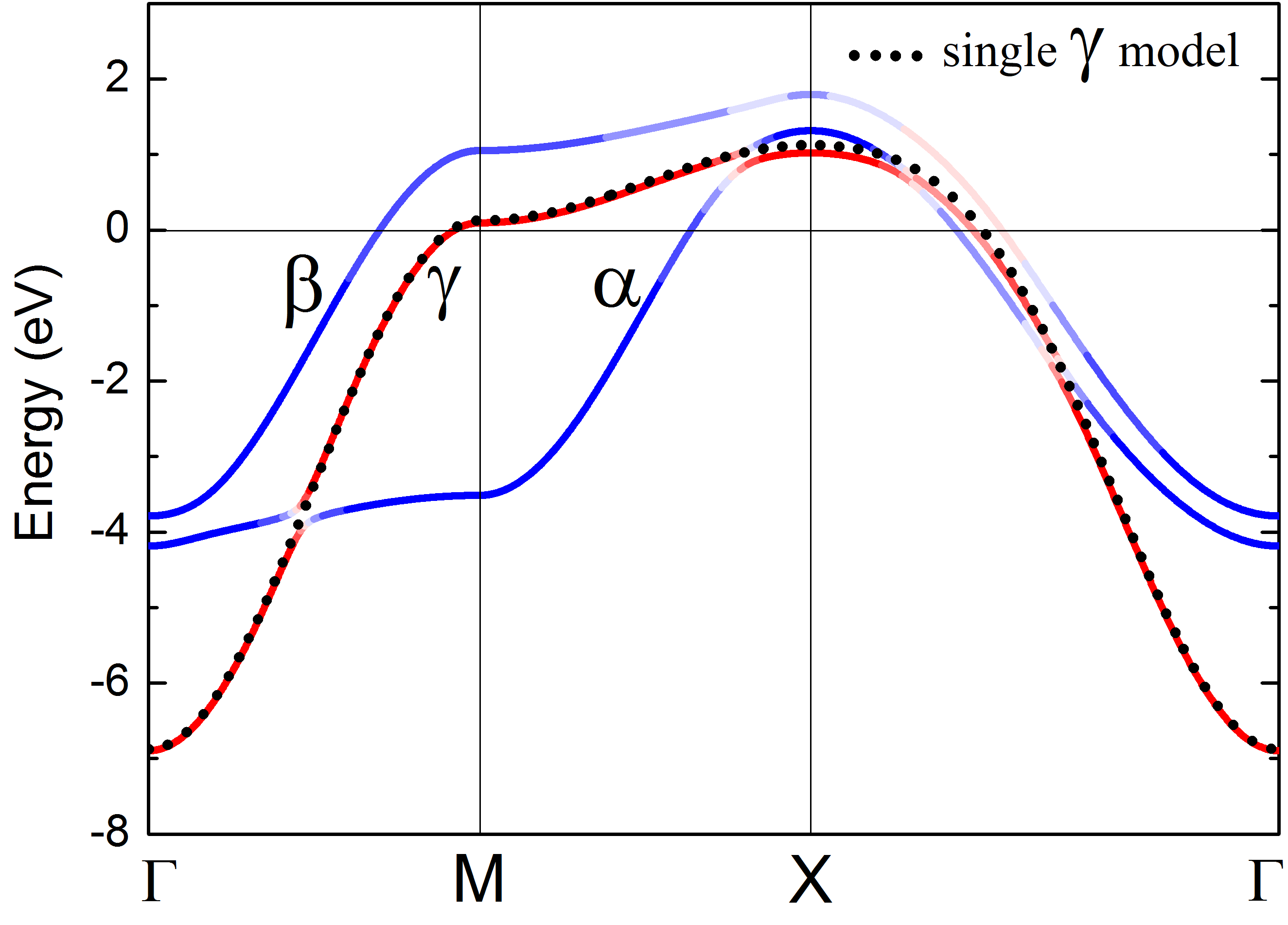}
\caption{Comparison of the $\gamma$ band structure in the single-band and three-band model. The deep blue and deep red portions of the solid lines correspond to the states with predominantly $d_{xz/yz}$ and $d_{xy}$ characters, respectively. We have assumed $t=1.25$eV.}
\label{fig:Bands}
\end{figure}

\begin{figure}
\includegraphics[width=8cm]{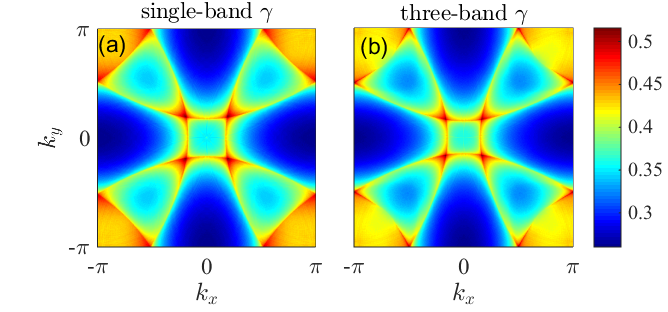}
\caption{$\gamma$-band bare susceptibility in: (a) the single-$\gamma$-band model and (b) the three-band model. The single-band model also uses $\mu_1=1.1t$, with its dispersion given in the text. The three-band model is described in the main text, and we have used $(t^{\prime\prime},\eta)=(0.05,0.15)t$. }
\label{fig:ChiGamma}
\end{figure}

In this section we perform calculations with an effective one-band model to make comparison with the three-band results. We adopt the same tight-binding dispersion as in the main text: $\xi_{\bs k}= -2t^\prime (\cos k_x + \cos k_y) - 4t^{\prime\prime\prime} \cos k_x \cos k_y -\mu_1$ with $(t^\prime,t^{\prime\prime\prime})=(0.8,0.3)t$, wherein the van Hove physics is retained. The corresponding band structure very well reproduces that of the $\gamma$-band in the three-band model except around the band mixing points (Fig \ref{fig:Bands}). For the interactions, only onsite Coulomb repulsion is included. Near van Hove filling, this band structure is known to generate enhanced spin fluctuations at a small momentum as well as at large wavevectors surrounding $(\pi,\pi)$. As in Fig \ref{fig:ChiGamma}, the single-band model almost exactly reproduces the bare susceptibility of the three-band model, suggesting limited influence of the high energy band crossing and SOC on the low-energy spin dynamics. At $\mu_1=1.1t$ (same as what was used in the main text), $d_{x^2-y^2}$-pairing dominates in all range of $U$ where RPA is valid, and $p$- and $s$-waves are subleading. The respective gap functions are shown in Fig 3 (d-f) in the main text. However, with slightly higher filling fraction $\mu=1.15t$, $p$-wave becomes more favorable at small $U$, with $s$-wave at intermediate and $d$-wave at the strongest $U$. Their gap functions are qualitatively the same as those shown in Fig 3 (d-f) in the main text.

\begin{thebibliography}{99}
\bibitem{footnote} Note that in our RPA calculations the peak value of the bare susceptibility components (S9) typically does not exceed $\sim 0.5$ for $0.1<U/t < \mathcal{O}(1)$.
\end{thebibliography}
\end{document}